\documentclass[conference]{IEEEtran}
\usepackage{subfigure}
\usepackage{lineno,hyperref}
\usepackage[table,xcdraw]{xcolor}
\usepackage{amsmath}
\usepackage{cases}
\usepackage{graphicx}
\usepackage[margin=1in]{geometry}
\usepackage{color}
\usepackage{array}
\usepackage{multirow}
\newcolumntype{P}[1]{>{\centering\arraybackslash}p{#1}}

\begin{document}
\title{A note on load balancing in DC microgrids}
\author{Shravan Mohan$^1$ and Bharath Bhikkaji$^2$ \\ $^1$Mantri Residency, Bannerghatta Road, Bangalore.  \\$^2$Indian Institute of Technology Madras, Chennai.
\thanks{}
}
\newgeometry{top=1in,bottom=0.75in,right=0.75in,left=0.75in}
\maketitle

\begin{abstract}
A problem of load balancing in isolated DC microgrids is considered in this paper. Here, a DC load is fed by multiple heterogenous DC sources, each of which is connected to the load via a boost converter. The gains of the DCC's provide for a means to control the division of load current amongst the DC sources. The primary objective of the control scheme is to minimise the total losses in the network, while maintaining the output voltage within a desired range, serving the load current demand and adhering to VI-characteristics of the power sources. Under assumptions of concavity/monotonocity/piece-wise-linearity of the VI-characteristics, the problem is solved using a convex relaxation. It is shown that the solution to the relaxed problem is tight. Thus, the resulting algorithm is guaranteed to reach global optimality in a numerically efficient manner. Simulations are provided for corroboration.
\end{abstract}

\section{Introduction}
\begin{figure}
    \centering
    \includegraphics[width=3.5in]{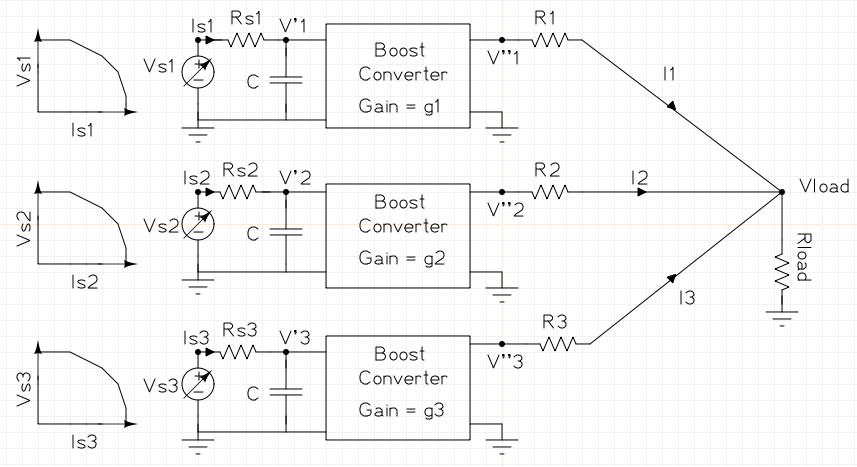}
    \caption{A schematic of the DC microgrid showing three DC sources, all together feeding the load. Each one has a different VI-characteristic as shown in the schematic and a different internal resistance. The load is modelled as a resistance. Also, each source is connected to the load via a boost converter. The connecting cables from the output of the DCC's to the load offer resistances as shown. The gains of the DCC's are the control inputs for a desirable load balance.}\vspace{-0.5cm}
    \label{fig:main_circuit}
\end{figure}

\noindent Consider the circuit shown in Figure \ref{fig:main_circuit}. The voltage sources with the VI-characteristic curves (output voltage vs current drawn) drawn beside those represent the heterogeneous power sources. Heterogeneity here is with respect to the VI characteristics. It is further assumed that all the VI characteristic curves are positive, piece-wise linear (PWL), concave and non-increasing (\textbf{A1}). Such an assumption, as shall be shown, aids analysis. It is also not an impractical assumption since many power sources show such behaviour and a concave curve can be well-approximated as a piece-wise linear curve. The internal resistances of these sources, $Rs_1$, $Rs_2$, $\cdots$, $Rs_k$, as also shown in Figure \ref{fig:main_circuit}, are considered separately and are not a part of the VI-characteristics (\textbf{A2}).  Each of these sources is connected to the load via a DC-DC converter (DCC). A large capacitive filter is placed at the input of the DCC's so as to prevent current ripple in the power sources (\textbf{A3}). It is assumed that the DCC's  in this problem are of the boost class (\textbf{A4}). The output of these DCC's are connected to the load via cable wires which offer their own resistances, $R_1$, $R_2$, $\cdots$, $R_k$. The load is assumed to be resistive (\textbf{A5}), which needs the voltage across it to be within a desired range. It is also assumed that the minimum of this range is greater than the open circuit voltages of all power sources (\textbf{A6}).\\\\
The DCC gains are the control inputs in this problem. It is mandated that the DCC's are operated in the Continuous Conduction Mode (CCM), by ensuring a minimum average current through the inductor (\textbf{A7}). The objective here is to set the gains so as to minimize the total losses in the system. The losses considered here are the resistive losses in the cables and the internal resistances, and in the DCC themselves. Note that the losses here refer to the losses at steady state; the losses during transients are not considered. In addition, while formulating an optimization problem, the fact that the DCC's do not inject any power into the system, must also be included. \\\\
Such scenarios commonly arise in systems fed by multiple renewable sources, such as solar panels, fuel cells, lithium-ion/lead-acid batteries and windmills. The variations in solar power, the fuel cell state and wind direction, or the variations in manufacturing process itself, leads to variations in the VI-characteristic curves of these devices. This makes a case for the requirement of an appropriate control strategy to balance the load current and regulate the load voltage. Since such networks might involve long cables, there is also a strong need to minimize the resistive losses in them. In addition, there are the losses of the DCC's themselves, which need to be minimized in order for an efficient thermal management. The losses in the DCC's are typically divided into conduction and switching losses. It will however be assumed that the conduction losses due to the current ripple component and the losses in the capacitors are negligible $\left(\textbf{A8}\right)$. \\\\
There has been considerable interest in this field. The authors of \cite{augustine2014adaptive} consider a similar load balancing in an isolated Dc microgrid and propose a metric termed as droop index, which measures the performance of the network. The performance is a function of the current difference between branches and the cable losses at the output side of the DCC's. The authors of \cite{prabhakaran2017novel} propose a decentralized control scheme to regulate the load voltage. The scheme depends only the output voltage and output current of a branch, thereby not requiring a communication overhead between the DCC's and a central controller. The authors of \cite{jabr2021mixed} present an optimization methodology for determining droop resistances, chosen from a discrete set of resistance, for various generation units in network so to minimize the resistive losses in the network. The focus of all the references mentioned here is to introduce a virtual resistance (called droop resistance) to improve current sharing amongst the loads, while maintaining voltage regulation. To the best of the authors' knowledge, an approach similar to the one presented for load balancing in DC-microgrids considering sources, which considers (i) varied concave non-increasing PWL VI-characteristic curves, (ii) the losses in DCC's, the internal resistances and the cables, (iii) while maintaining an output voltage within a desired range and (iv) catering to a current demand, has not been discussed in literature. 
\cite{erickson2007fundamentals}
\cite{boyd2004convex},  \cite{mohan2007power}, \cite{holmes2003pulse}, \cite{augustine2014adaptive}, \cite{prabhakaran2017novel} , \cite{diamond2016cvxpy}\\\\
\section{The Optimization Problem}
\noindent The following notations describe the optimization variables in a network with $N$ branches.
\begin{itemize}
    \item $\left\{Vs_{1}, Vs_{2}, \cdots, Vs_{N}\right\}$ denote the average source voltages at steady state.
    
    \item $\left\{Is_{1}, Is_{2}, \cdots, Is_{N}\right\}$  denote the average source currents at steady state, while $\left\{is_{1}(t), is_{2}(t), \cdots, is_{N}(t)\right\}$ represent the instantaneous source currents.
    
    \item  $\left\{V'_1, V'_2, \cdots, V'_N\right\}$ denote the average DCC's' average input voltages at steady state.
    
    \item  $\left\{V''_1, V''_2, \cdots, V''_N\right\}$ denote the average DCC's' average output voltages at steady state.
    
    \item  $\left\{I_{1}, I_{2}, \cdots, I_{N}\right\}$ denote the average DCC's' average output currents at steady state.
    
    \item $\left\{id_{1}(t), id_{2}(t), \cdots, id_{N}(t)\right\}$ represent the instantaneous diode currents, and $\left\{im_{1}(t), im_{2}(t), \cdots, im_{N}(t)\right\}$ represent the instantaneous MOSFET currents.
    
    \item $\left\{vd_{1}(t), vd_{2}(t), \cdots, vd_{N}(t)\right\}$ represent the instantaneous voltage at the anode of the diode.
\end{itemize}
In the following analysis, the time $(t)$ will be dropped for ease of exposition. Also, bold lettered variables will denote the optimal values. The following are known parameters of the circuit components.
\begin{itemize}
    \item $\displaystyle f_k(x) = \min_{i\in \left[1,P_k\right]} \left\{\beta_{k,i}x + \gamma_{k,i}\right\}$,$~\beta_{k,i}<0 ~\forall~ k,i$ (see \textbf{A1}).
    
    \item  $\left\{Rs_1, Rs_2, \cdots, Rs_N\right\}$ and $\left\{R_1, R_2, \cdots, R_N\right\}$, resistances connecting the sources to the DCC's (see \textbf{A2}) and the DCC's to the load, respectively.
    
    \item  $R_{\rm load}$, the load resistance.
    
    \item  $\left\{R_{L,1}, R_{L,2},\cdots, R_{L,N}\right\}$, the DC resistances of the inductors used in the DCC's.
    
    \item  $\left[V_{L, \min}, V_{L, \max}\right]$, desired average output voltage range.
    
    \item  $\left\{I_{1, \min}, I_{2, \min},\cdots,I_{N, \min} \right\}$, minimum average output currents to ensure CCM.
    
    \item  $\left\{g_{1,\max}, g_{2,\max}, \cdots, g_{N,\max}\right\}$,  upper bounds on the DCC gains. The lower bounds are assumed to be one (see \textbf{A4}). 
    
    \item  $\left\{\lambda_1, \lambda_2, \cdots, \lambda_N\right\}$ and $\left\{\mu_1, \mu_2, \cdots, \mu_N\right\}$, user-defined weights.
    
    \item  $\left\{R_{M,1}, R_{M,2},\cdots, R_{M,N}\right\}$, ON state drain-source resistances of the MOSFETs used in the DCC's.
    
    \item  $\left\{V_{D,1}, V_{D,2},\cdots, V_{D,N}\right\}$, the forward bias voltages of diodes used in the DCC's.
    
     \item  $\left\{R_{D,1}, R_{D,2},\cdots, R_{D,N}\right\}$, the diode resistances.
    
    \item  $\left\{\alpha_{1}, \alpha_{2},\cdots, \alpha_{N}\right\}$, the multiplicative constant for deriving switching losses for the MOSFETs.  
\end{itemize}
\subsection{The Construct}
Consider the optimization problem mentioned below:
\begin{align}
     \displaystyle &\min~\sum_{k=1}^N \lambda_k \left\{Is_{k}^2\left(Rs_k+R_{L,k}\right) + \left(im_k\right)_{\rm rms}^2R_{M,k} \cdots \right.\nonumber\\
     &\left...  + \left(id_k\right)_{\rm rms}^2R_{D,k} + V_{D,k}I_k + I_{k}^2R_k + \alpha_{ k}\left(vd_k id_k\right)_{\rm peak} \right\}  \\
         &\mbox{subject to}\nonumber \\
         &~~~~~~V'_k = Vs_k - Is_kRs_k,~\forall k,\\
         &~~~~~~V''_k = V_{\rm load} + I_kR_k,~\forall k,\\
         &~~~~~~\sum_{k=1}^N I_k = \frac{V_{\rm load}}{R_{\rm load}},\\
         &~~~~~~V_{\rm load,\min} \leq V_{\rm load} \leq V_{l,\max},\\
         &~~~~~~V'_k\leq V''_k \leq g_{k,\max}V'_k, ~\forall k,\\
         &~~~~~~Vs_k = f_k\left(Is_k\right) ,~\forall k,\\
         &~~~~~~Vs_kIs_k = \left\{Is_{k}^2\left(Rs_k+R_{L,k}\right) + \left(im_k\right)_{\rm rms}^2R_{M,k} \cdots \right.\nonumber\\
        &~~~~~~~~~\left.\cdots  + \left(id_k\right)_{\rm rms}^2R_{D,k} + I_{k}^2R_k + V_{D,k}I_k\cdots\right. \nonumber \\
        &~~~~~~~~~\left.\cdots + \alpha_{k}\left(vd_kid_k\right)_{\rm peak} +V_{\rm load} I_k\right\} ~\forall k.\\
         &~~~~~~Is_k\geq Is_{k,\min}, ~~Vs_k,Is_k,V'_k,V''_k \geq 0, \forall k.
         \label{eqn:orig_opt_problem}
\end{align}
Modifications to this problem, which will be discussed in the following subsections,  will lead to the main algorithm of this paper shown in Figure \ref{fig:mainalgo}. \\
\subsubsection{The Cost Function}
The objective function is the weighted sum total of losses in the cables, the internal resistances of sources and the DCC's. These weights can be regarded as the cost of the losses in a particular branch, based on the type of power source used. With large input and output capacitors (see \textbf{A3}) in the DCC's, the ripple in the input source current and the output current is negligible. Thus, the resistive losses in the cables and the internal resistances of the $k^{\rm th}$ branch is 
\begin{align}
    I_k^2R_k + Is^2_k Rs_k.
    \label{eqn:internalreistancecablelosses}
\end{align}
The loss in a DCC is composed of two parts: (i) the conduction losses in the inductor, the switch and the diode and (ii) the switching loss in the switch. Since the current ripple in the inductor is very low (see \textbf{A8}) as compared to the average value, the conduction loss in inductor of the  $k^{\rm th}$ converter is approximately 
\begin{align}
   Is_k^2R_{L,k} 
   \label{eqn:inductorconductionloss}
\end{align}
Now, notice that in the boost converter, the instantaneous current through the inductor is the sum total of the instantaneous switch current and diode current. Thus
\begin{align}
    is_k = im_k + id_{\rm k} \Longrightarrow Is_k = <im_k> + I_k,
    \label{eqn:kclatdrainMOSFET}
\end{align}
where $<.>$ denotes the time average over one time period. Also note that the instantaneous switch current and the diode current in a boost converter are orthogonal. That is,
\begin{align}
    im_{k} id_k = \left(is_k - id_k\right)id_k = 0.
    \label{eqn:currentorthogonality}
\end{align}
With these observations, the conduction loss in the switch of the $k^{\rm th}$ converter is approximated as 
\begin{align}
    \left(im_k\right)_{\rm rms}^2 R_{M,k} =& \frac{1}{T}\int_{0}^T \left(is_k-id_k\right)^2(t) R_{M,k} dt \nonumber \\=& \frac{1}{T}\int_{0}^T \left(is_k\left(is_k-id_k\right)\right)(t) R_{M,k} dt \nonumber \\\approx  Is_k(Is_k-I_k)R_{M,k} =& Is_k^2R_{M,k} - Is_kI_kR_{M,k}.
    \label{eqn:mosfetconductionloss}
\end{align}
Using a similar logic, the conduction loss in the diode of the $k^{\rm th}$ converter is 
\begin{align}
    V_{D,k}I_k + \left(id_k\right)_{\rm rms}^2R_{D,k} =  V_{D,k}I_k + Is_kI_kR_{D,k}.
    \label{eqn:diodeconductionloss}
\end{align}
The switching loss in the MOSFET of the $k^{\rm th}$ converter is \cite{rohm_app_note, erickson2007fundamentals}
\begin{align}
    &\alpha_{k} \left(vd_{k}id_k\right)_{\rm peak} = \alpha_{k}\left(vd_{k}\right)_{\rm peak}\left(id_k\right)_{\rm \tiny peak} = \alpha_{k}\left(V_{\rm load}+I_kR_k\right... \nonumber\\& \cdots \left.V_{D,k}+Is_kR_{D,k}\right)Is_k = \alpha_{k}\left(V_{\rm load}+V_{D,k}\right)Is_k..\nonumber\\ &\cdots+ \alpha_{k}Is_kI_kR_k + \alpha_{k}Is_k^2R_{D,k}.
    \label{eqn:mosfetswitchingloss}
\end{align}
Therefore, the cost function is 
\begin{align}
    \displaystyle  &\sum_{k=0}^N\lambda_k\left\{Is_k^2\left(Rs_k + R_{L,k} + R_{M,k} + \alpha_{k}R_{D,k}\right) ..\right. \nonumber \\ 
    &\left.\cdots + I_k^2R_k + Is_kI_k \left(\alpha_{\rm k}R_k - \left(R_{M,k} - R_{D,k}\right)\right)\cdots\right. \nonumber\\&\left.\cdots + \alpha_{\rm k} V_{\rm load}I_k + V_{D,k}\left(I_k+\alpha_k Is_k\right)\right\}. 
    \label{eqn:costfundraft1}
\end{align}
The above can be modified further as
\begin{align}
    \displaystyle &\sum_{k=0}^N\lambda_kQ_k =\sum_{k=0}^N\lambda_k \left\{Is_k^2\left(R_{\rm eff, 1} -R_{\rm eff, 2}\right)\cdots \right. \nonumber \\&\cdots + \left(I_k + Is_k\alpha_{\rm k}\right)^2\frac{R_k}{2}  + \left(Is_k - sI_k\right)^2\frac{\left|R_{M,k}-R_{D,k}\right|}{2} \cdots\nonumber \\&\left.\cdots + \alpha_{\rm k} V_{\rm load}I_k  + V_{D,k}\left(I_k+\alpha_kIs_k\right)+ I_k^2R_{\rm eff, 3} 
     \right\} ,
     \label{eqn:costfunfinal}
\end{align}
where $s = \mbox{sign}\left(R_{M,k}-R_{D,k}\right)$, 
\begin{align}
    R_{\rm eff, 1} = \left(Rs_k + R_{L,k} +R_{M,k} + \alpha_{\rm k}R_{D,k}\right),
    \label{eqn:reff1}
\end{align}
\begin{align}
    R_{\rm eff, 2} = \left( \frac{\left|R_{M,k}-R_{D,k}\right|}{2} + \frac{\alpha_{\rm k}^2 R_k}{2}\right),
    \label{eqn:reff2}
\end{align}
and
\begin{align}
    R_{\rm eff, 3} = \left(\frac{R_k-\left|R_{M,k}-R_{D,k}\right|}{2}\right).
    \label{eqn:reff3}
\end{align}
The expression is convex iff 
\begin{equation}
   R_{\rm eff, 1}  \geq R_{\rm eff, 2} ~\mbox{and}~
    R_{\rm eff, 3} \geq 0.
    \label{eqn:conditions}
\end{equation}
These are not an impractical conditions for the following reason. The quantity $\alpha_{\rm k}$ is approximately
\begin{align}
    \alpha_{\rm k} = \frac{1}{2} \left(\tau_{\rm ON} + \tau_{\rm OFF}\right) f_s,
    \label{eqn:alphaexpression}
\end{align}
where $\tau_{\rm ON}$ and $\tau_{\rm OFF}$ are ON and OFF transition times for a MOSFET. For typical values $f_s = 100$ KHz, and $\tau_{\rm ON}+\tau_{\rm OFF} = 100$ ns, one obtains $\alpha_{ k} \approx 0.005$. With a typical large inductor resistance of $1\Omega$, a source resistance of $500m\Omega$, MOSFET ON-state resistance of $200m\Omega$, diode resistance of $30m\Omega$ and a low cable resistance of $1\Omega$ the bounds reads 
\begin{align}
    1.7 \geq 0.115 \mbox{~and~} 1 \geq 0.115.
    \label{eqn:conditionsexample}
\end{align}
Thus, the two conditions are satisfied with healthy margins, thereby resulting in a convex cost function in practical cases. \\\\
There might also be an interest to add to the cost a weighted combination of the absolute values of circulating currents in the network. Circulating currents arise due to the difference between the output voltages of the different DCC's. Mathematically, the circulating currents are a linear transform of the output voltages of the DCC's, and a weighted absolute sum (by weights $\left\{\mu_1, \mu_2,\cdots,\mu_N\right\}$) of these adds a convex term to the cost function. That is, the total circulating current from $k^{\rm th}$ converter is given as \cite{augustine2014adaptive}
\begin{align}
    Ic_k = \sum_{j\neq k}\frac{V''_k-V''_j}{(R_k+R_j)},
    \label{eqn:circulatingcurrent}
\end{align}
and thus the additive cost would be 
\begin{align}
    \sum_{k=1}^N\mu_k|Ic_k| = \sum_{k=1}^N\mu_k\left|\sum_{j\neq k}\frac{V''_k-V''_j}{(R_k+R_j)}\right|,
    \label{eqn:additive cost}
\end{align}
It will be shown that with this additive term in the cost function, the qualitative aspects of the solution remain unchanged. \\
\subsubsection{The Constraints}
There are in all eight sets of constraints in the optimization problem \eqref{eqn:orig_opt_problem}. The first set of constraints are the Kirchoff's Voltage Laws (KVL) applied to the input side of the DCC's. The second set is the KVL applied to the output side of the DCC's (see \textbf{A5}). The third constraint is the Kirchoff's Current Law applied to load node. The fourth constraint bounds the output voltage within the desired range. The fifth set of constraints bounds the gains of the DCC's within practical ranges (see \textbf{A4}). For example, the following expression gives a conservative upper bound on the the gain of the $k^{\rm th}$ boost converter (ignoring switching loss) \cite{erickson2007fundamentals},
\begin{align}
    \frac {\left(1-D'\frac{V_{D,k}}{V_{in}}\right)D'\left(R_k+R_{\rm load}\right)} {(D')^2\left(R_k+R_{\rm load}\right) + D'R_{D,k}+DR_{M,k}+R_{L,k}},
    \label{eqn:gainexpression}
\end{align}
where $D'=1-D$. Note that gain depends on the input voltage to the DCC. Suppose now that the input voltage is constrained to be at least 20 times the forward bias of the diode. That is
\begin{align}
    V'_k \geq 20*V_{D,k}.
    \label{eqn:inputvoltagelowerbound}
\end{align}
Then a conservative maximum on the gain would be
\begin{align}
    \max_{D}\left\{\frac {0.95 D'\left(R_k+R_{\rm load}\right)} {(D')^2\left(R_k+R_{\rm load}\right) + D'R_{D,k}+DR_{M,k}+R_{L,k}}\right\}.
    \label{eqn:conservativegainbound}
\end{align}
The sixth set of constraints ensures that the VI-characteristics of the power sources are adhered to (see \textbf{A1}). The seventh set of constraints ensures that the total power input to the DCC is equal its output power plus the DCC losses \cite{erickson2007fundamentals}. This is equivalent to stating that the DCC's do not inject power into the network. To derive the functional form of the constraint, note that
\begin{align}
    V'_kIs_k &= V''_kI_k + Q_k\nonumber\\
    \Rightarrow \left(Vs_k-Is_kRs_k\right)Is_k &= \left(V_{\rm load}+I_kR_k\right)I_k + Q_k\nonumber \\
    \Rightarrow Vs_kIs_k &= Is_k^2Rs_k + Q_k + I_k^2R_k + V_{\rm load}I_k\nonumber.
\end{align}
Substituting the expression for $Q_k$, as discussed earlier, one obtains the seventh set of  constraints.\\\\
The last set of constraints, requiring the voltage and current opt-variables to be non-negative, ensures that power is only drawn from the power sources. Now suppose that the following lower bounds hold
\begin{align}
    Is_k \geq 5 \frac{\max\{Vs_k\}}{f_s L_k}, ~\forall k.     
    \label{eqn:sourcecurrentCCM}
\end{align} 
It is immediate that converters would operate only in CCM as desired, as the average current is  larger than the maximum deviation that can occur (see \textbf{A7}). Moreover, the RMS of a waveform with an average of $Is_k$ and a linear deviation of at most $40\%$ is given by \cite{erickson2007fundamentals}
\begin{align}
    \left(is_k\right)_{\rm rms}^2 = Is_k^2\left(1 + \frac{0.4^2}{12}\right) \approx Is_k^2,
    \label{eqn:rmsapprox}
\end{align}
so that the conduction losses in the inductors are approximated well by considering only the average current passing through those. Finally,  the optimal gains can be calculated as the ratio of the optimal $V''_k$'s to $V'_k$'s, and the optimal duty ratios can be determined using \cite{erickson2007fundamentals}
\begin{align}
    \boldsymbol{I_k} \approx \left(1-\boldsymbol{D}\right)\boldsymbol{Is_k}.
    \label{eqn:dutyratioexpression}
\end{align}
It may be noted that the duty ratio corresponding to optimal gains need not be perfectly attainable in practice due to the limit on the counter frequency. In that case, it is assumed that the duty cycle in implementation will be the best approximation to the optimal duty ratio. 

\begin{figure*}[t]
\centering
\fbox{\begin{minipage}{7in}
\begin{enumerate}
    \item Set $V_{\rm load}$ to $V_{\rm load,\min}$. Ensure that given circuit parameters adhere to \eqref{eqn:conditions}.
    \item {Solve the following:
    \vspace{-0.1cm}
    \begin{align}
          &\hspace{-0.7cm}\min\sum_{k=0}^N\lambda_k\left\{Is_k^2\left(R_{\rm eff, 1} - R_{\rm eff, 2}\right)\right. + \left(I_k + Is_k\alpha_{\rm k}\right)^2\frac{R_k}{2} + \left(Is_k - sI_k\right)^2\frac{\left|R_{M,k}-R_{D,k}\right|}{2}+ \alpha_{\rm k} V_{\rm load}I_k + V_{D,k}I_k \left. + I_k^2R_{\rm eff, 3} \right\}\nonumber\\
             &\hspace{-0.7cm}\mbox{subject to}\nonumber \\
             &\hspace{-0.7cm}~~~V'_k = Vs_{k} - Is_kRs_k~\forall k, 
             ~~V''_k = V_{\rm load} + I_kR_k~\forall k,
             ~~\sum_{k=1}^mI_k = \frac{V_{\rm load}}{R_{\rm load}},
             ~~V'_k\leq V''_k \leq g_{k,\max}V'_k ~~\forall k,
             ~~V_k \leq f_k\left(Is_k\right) ~\forall k,\nonumber \\
             &\hspace{-0.5cm}\min_{j\in [1,P_k]} \left\{ \beta_{k,j}Is^2_k + \gamma_{k,j}Is_k\right\} \geq \left.Is_k^2\left(R_{\rm eff, 1} - R_{\rm eff, 2}\right)\right. \left(I_k + Is_k\alpha_{\rm k}\right)^2\frac{R_k}{2} + \left(Is_k - sI_k\right)^2\frac{\left|R_{M,k}-R_{D,k}\right|}{2} \cdots \nonumber \\&\hspace{1cm}\cdots + \alpha_{\rm k} V_{\rm load}I_k + V_{D,k}I_k \left. + I_k^2R_{\rm eff, 3} + V_{\rm load}I_k ~\forall k
     \right. \nonumber,\\
         &\hspace{-0.2cm}I_k\geq I_{k,\min}~\forall k, ~~V_k,Is_k,V'_k,V''_k \geq 0~\forall k.\nonumber 
    \end{align}
    }
    \item If feasible, and for some $k$, $\boldsymbol{V_k}<f_k\left(\boldsymbol{Is_k}\right)$, then set $V_k$ to $f_k\left(\boldsymbol{Is_k}\right)$, set $V'_k$ to $\boldsymbol{V'_k} + f_k\left(\boldsymbol{Is_k}\right) - \boldsymbol{V_k}$.
\end{enumerate}
\end{minipage}}
\caption{The Main Optimization Algorithm}
\label{fig:mainalgo}
\end{figure*}

\begin{table*}[!]
\vspace{-0.4cm}
\centering
\caption{Case(i): Parameters for the concave power sources}
\vspace{-0.2cm}
\resizebox{6.9in}{!}{\begin{tabular}{|c|c|c|c|c|c|c|c|c|c|c|}
\hline
Br. No. & $\beta_1, \gamma_1$                                         & $\beta_2, \gamma_2$                                         & $\beta_3, \gamma_3$                                        & $\beta_4, \gamma_4$                                         & $\beta_5, \gamma_5$                                         & $\beta_6, \gamma_6$                                         & $\beta_7, \gamma_7$                                         & $\beta_8, \gamma_8$                                         & $\beta_9, \gamma_9$                                         & $\beta_{10}, \gamma_{10}$                                  \\ \hline
1      & \begin{tabular}[c]{@{}c@{}}-0.4483, \\ 42.2458\end{tabular} & \begin{tabular}[c]{@{}c@{}}-0.7394, \\ 42.8602\end{tabular} & \begin{tabular}[c]{@{}c@{}}-0.848, \\ 43.3191\end{tabular} & \begin{tabular}[c]{@{}c@{}}-1.2129, \\ 45.6297\end{tabular} & \begin{tabular}[c]{@{}c@{}}-2.0441, \\ 52.6495\end{tabular} & \begin{tabular}[c]{@{}c@{}}-2.8258,\\  60.8998\end{tabular} & \begin{tabular}[c]{@{}c@{}}-3.4599, \\ 68.932\end{tabular}  & \begin{tabular}[c]{@{}c@{}}-3.806,\\  74.0474\end{tabular}  & \begin{tabular}[c]{@{}c@{}}-4.6268,\\  87.9088\end{tabular} & \begin{tabular}[c]{@{}c@{}}-4.8629,\\ 92.3943\end{tabular} \\ \hline
2      & \begin{tabular}[c]{@{}c@{}}-0.1616, \\ 35.5302\end{tabular} & \begin{tabular}[c]{@{}c@{}}-0.3842,\\  35.9755\end{tabular} & \begin{tabular}[c]{@{}c@{}}-0.911, \\ 38.0826\end{tabular} & \begin{tabular}[c]{@{}c@{}}-1.7442, \\ 43.0815\end{tabular} & \begin{tabular}[c]{@{}c@{}}-1.982, \\ 44.9842\end{tabular}  & \begin{tabular}[c]{@{}c@{}}-2.4405, \\ 49.5688\end{tabular} & \begin{tabular}[c]{@{}c@{}}-2.4844, \\ 50.0966\end{tabular} & \begin{tabular}[c]{@{}c@{}}-3.4, \\ 62.9142\end{tabular}    & \begin{tabular}[c]{@{}c@{}}-4.2573, \\ 76.631\end{tabular}  & \begin{tabular}[c]{@{}c@{}}-4.9458,\\  89.024\end{tabular} \\ \hline
3      & \begin{tabular}[c]{@{}c@{}}-0.8106, \\ 28.1456\end{tabular} & \begin{tabular}[c]{@{}c@{}}-0.8138, \\ 28.1517\end{tabular} & \begin{tabular}[c]{@{}c@{}}-0.858, \\ 28.3187\end{tabular} & \begin{tabular}[c]{@{}c@{}}-1.0573, \\ 29.4479\end{tabular} & \begin{tabular}[c]{@{}c@{}}-1.1274, \\ 29.9775\end{tabular} & \begin{tabular}[c]{@{}c@{}}-1.8309, \\ 36.6217\end{tabular} & \begin{tabular}[c]{@{}c@{}}-2.0942, \\ 39.6052\end{tabular} & \begin{tabular}[c]{@{}c@{}}-2.7073, \\ 47.7127\end{tabular} & \begin{tabular}[c]{@{}c@{}}-3.6009,\\  61.2161\end{tabular} & \begin{tabular}[c]{@{}c@{}}-4.3478,\\  73.912\end{tabular} \\ \hline
\end{tabular}}
\label{tab:caseisourceparams}
\end{table*}

\begin{table*}[t]
\vspace{-0.4cm}
\centering
\caption{Case(i): Other circuit parameters and results, $V_{\rm load}=50,~R_{\rm load}=5$}
\vspace{-0.2cm}
\resizebox{6.9in}{!}{\begin{tabular}{|c|c|c|c|c|c|c|c|c|c|c|c|c|c|c|c|c|c|}
\hline
Br. No.  & $Rs$ & $R_{L}$ & $R_M$ & $\alpha$ & $V_D$ & $R_D$ & $I_{\min}$ & $V'_{\min}$ & $g_{\max}$ & $\lambda$ & $\mu$ & {\color[HTML]{6200C9} $Is$}   & {\color[HTML]{6200C9} $V'$}    & {\color[HTML]{6200C9} $V''$}   & {\color[HTML]{6200C9} $I$}    \\ \hline
1          & 0.5  & 0.04  & 0.019 & 0.0026   & 0.6   & 0.014 & 0.6613     & 12          & 4.4755     & 1         & 1     & {\color[HTML]{6200C9} \begin{tabular}[c]{@{}c@{}}6.9648,\\6.9561\end{tabular}}   & {\color[HTML]{6200C9} \begin{tabular}[c]{@{}c@{}}33.6996, \\33.699\end{tabular}}   & {\color[HTML]{6200C9} \begin{tabular}[c]{@{}c@{}}50.8974, \\50.894\end{tabular}} & {\color[HTML]{6200C9} \begin{tabular}[c]{@{}c@{}}4.4870,\\4.4865\end{tabular}} \\ \hline
2         & 0.4  & 0.053 & 0.019 & 0.0026   & 0.6   & 0.014 & 0.3003     & 12          & 4.0702     & 1.5       & 1     & {\color[HTML]{6200C9} \begin{tabular}[c]{@{}c@{}}5.5893,\\5.5905\end{tabular}} & {\color[HTML]{6200C9}\begin{tabular}[c]{@{}c@{}} 30.7549, \\30.753\end{tabular}} & {\color[HTML]{6200C9} \begin{tabular}[c]{@{}c@{}}50.8216,\\50.818\end{tabular}} & {\color[HTML]{6200C9}\begin{tabular}[c]{@{}c@{}} 3.2864,\\3.2865\end{tabular}} \\ \hline
3         & 0.45 & 0.053 & 0.019 & 0.0026   & 0.6   & 0014  & 0.2012     & 12          & 4.0627    & 1         & 1     & {\color[HTML]{6200C9} \begin{tabular}[c]{@{}c@{}}5.5357,\\5.5364\end{tabular}}  & {\color[HTML]{6200C9}\begin{tabular}[c]{@{}c@{}} 21.0780,\\21.077\end{tabular}} & {\color[HTML]{6200C9}\begin{tabular}[c]{@{}c@{}} 50.5120,\\50.509\end{tabular}} & {\color[HTML]{6200C9}\begin{tabular}[c]{@{}c@{}} 2.2264,\\2.2263\end{tabular}} \\ \hline
\end{tabular}}
\label{tab:caseiparamsandresults}
\end{table*}

\begin{table*}[t]
\vspace{-0.4cm}
\centering
\caption{Case (ii, a): Circuit parameters and results (CVXPY, LTSpice), $V_{\rm load}=70,~R_{\rm load}=5, ~V_{\rm load} \mbox{~from LTSpice}~=71.137$}
\vspace{-0.2cm}
\resizebox{6.9in}{!}{\begin{tabular}{|c|c|c|c|c|c|c|c|c|c|c|c|c|c|c|c|c|}
\hline
Br. No. & $V_s$ & $Rs$ & $R_{L}$ & $R_M$ & $\alpha$ & $V_D$ & $R_D$ & $I_{\min}$ & $V'_{\min}$ & $g_{\max}$ & $\lambda$ & $\mu$ & \color[HTML]{6200C9}$Is$& \color[HTML]{6200C9}$V'$& \color[HTML]{6200C9}$V''$& \color[HTML]{6200C9}$I$\\ \hline
1      & 50    & 0.5  & 0.04  & 0.019 & 0.002143   & 0.5418   & 0.0184 & 0.6643     & 10.8366          & 4.4576     & 1         & 1 & \color[HTML]{6200C9} \begin{tabular}[c]{@{}c@{}}8.8644,\\9.0461\end{tabular} & \color[HTML]{6200C9}\begin{tabular}[c]{@{}c@{}}45.5677,\\45.477\end{tabular}& \color[HTML]{6200C9}\begin{tabular}[c]{@{}c@{}}71.1108,\\72.251\end{tabular}& \color[HTML]{6200C9}\begin{tabular}[c]{@{}c@{}}5.5540,\\5.5644\end{tabular}  \\ \hline
2      & 45    & 0.4  & 0.053 & 0.019 & 0.002143   & 0.5418  & 0.0184 & 0.3447     & 10.8366          & 4.0555     & 1.5       & 1  & \color[HTML]{6200C9}\begin{tabular}[c]{@{}c@{}}7.2370,\\7.5524\end{tabular} & \color[HTML]{6200C9}\begin{tabular}[c]{@{}c@{}}42.1051,\\41.979\end{tabular}& \color[HTML]{6200C9}\begin{tabular}[c]{@{}c@{}}71.0471,\\72.209\end{tabular}& \color[HTML]{6200C9}\begin{tabular}[c]{@{}c@{}}4.1885,\\4.284\end{tabular}  \\ \hline
3      & 40    & 0.45 & 0.053 & 0.019 & 0.002143   & 0.5418   & 0.0184  & 0.2932     & 10.8366          & 4.0481     & 1         & 1   & \color[HTML]{6200C9}\begin{tabular}[c]{@{}c@{}}8.6130,\\9.0689\end{tabular} & \color[HTML]{6200C9}\begin{tabular}[c]{@{}c@{}}36.1241,\\35.919\end{tabular}& \color[HTML]{6200C9}\begin{tabular}[c]{@{}c@{}}70.9792,\\72.145\end{tabular}& \color[HTML]{6200C9}\begin{tabular}[c]{@{}c@{}}4.2574,\\4.3792\end{tabular} \\ \hline
\end{tabular}}
\label{tab:caseiia}
\vspace{-0.2cm}
\end{table*}

\begin{table}[t]
\vspace{-0.4cm}
\centering
\caption{Case(ii, a): Power loss numbers comparison}
\vspace{-0.2cm}
\resizebox{3.4in}{!}{\begin{tabular}{|c|c|c|c|c|}
\hline
Br. No. & \begin{tabular}[c]{@{}c@{}}MOSFET Loss\\ CVXPY\end{tabular} & \begin{tabular}[c]{@{}c@{}}MOSFET Loss\\ LTSpice\end{tabular} & \begin{tabular}[c]{@{}c@{}}Diode Loss\\ CVXPY\end{tabular} & \begin{tabular}[c]{@{}c@{}}Diode Loss\\ LTSpice\end{tabular} \\ \hline
1      & 1.9218 & 2.0574 & 3.9165  & 3.9462 \\ \hline
2      &  1.5315 & 1.6724 & 2.8280  & 2.9443 \\ \hline
3      & 2.0358 & 2.2614 & 2.9824  & 3.1069 \\ \hline
\end{tabular}}
\label{tab:caseiiapower}
\vspace{-0.4cm}
\end{table}

\begin{table*}[t]
\centering
\caption{Case (ii, b): Circuit parameters and results (CVXPY, LTSpice), $V_{\rm load}=70,~R_{\rm load}=5, ~V_{\rm load} \mbox{~from LTSpice}~=69.997$}
\vspace{-0.2cm}
\resizebox{6.9in}{!}{\begin{tabular}{|c|c|c|c|c|c|c|c|c|c|c|c|c|c|c|c|c|}
\hline
Br. No. & $V_s$ & $Rs$ & $R_{L}$ & $R_M$ & $\alpha$ & $V_D$ & $R_D$ & $I_{\min}$ & $V'_{\min}$ & $g_{\max}$ & $\lambda$ & $\mu$ & \color[HTML]{6200C9}$Is$& \color[HTML]{6200C9}$V'$& \color[HTML]{6200C9}$V''$& \color[HTML]{6200C9}$I$\\ \hline
1      & 50    & 0.5  & 0.04  & 0.019 & 0.002143   & 0.5418   & 0.0184 & 0.6643     & 10.8366          & 4.4576     & 1         & 1 & \color[HTML]{6200C9} \begin{tabular}[c]{@{}c@{}}8.8644,\\8.8648\end{tabular} & \color[HTML]{6200C9}\begin{tabular}[c]{@{}c@{}}45.5677,\\45.568\end{tabular}& \color[HTML]{6200C9}\begin{tabular}[c]{@{}c@{}}71.1108,\\71.108\end{tabular}& \color[HTML]{6200C9}\begin{tabular}[c]{@{}c@{}}5.5540,\\5.5536\end{tabular}  \\ \hline
2      & 45    & 0.4  & 0.053 & 0.019 & 0.002143   & 0.5418  & 0.0184 & 0.3447     & 10.8366          & 4.0555     & 1.5       & 1  & \color[HTML]{6200C9}\begin{tabular}[c]{@{}c@{}}7.2370,\\7.2405\end{tabular} & \color[HTML]{6200C9}\begin{tabular}[c]{@{}c@{}}42.1051,\\42.104\end{tabular}& \color[HTML]{6200C9}\begin{tabular}[c]{@{}c@{}}71.0471,\\71.045\end{tabular}& \color[HTML]{6200C9}\begin{tabular}[c]{@{}c@{}}4.1885,\\4.1899\end{tabular}  \\ \hline
3      & 40    & 0.45 & 0.053 & 0.019 & 0.002143   & 0.5418   & 0.0184  & 0.2932     & 10.8366          & 4.0481     & 1         & 1   & \color[HTML]{6200C9}\begin{tabular}[c]{@{}c@{}}8.6130,\\8.6128\end{tabular} & \color[HTML]{6200C9}\begin{tabular}[c]{@{}c@{}}36.1241,\\36.124\end{tabular}& \color[HTML]{6200C9}\begin{tabular}[c]{@{}c@{}}70.9792,\\70.976\end{tabular}& \color[HTML]{6200C9}\begin{tabular}[c]{@{}c@{}}4.2574,\\4.2561\end{tabular} \\ \hline
\end{tabular}}
\label{tab:caseiib}
\vspace{-0.2cm}
\end{table*}

\begin{table*}[t]
\vspace{-0.2cm}
\centering
\caption{Case (iii): Circuit parameters and results (CVXPY, LTSpice), $V_{\rm load}=70,~R_{\rm load}=5, ~V_{\rm load} \mbox{~from LTSpice}~=69.997$}
\vspace{-0.2cm}
\resizebox{6.9in}{!}{\begin{tabular}{|c|c|c|c|c|c|c|c|c|c|c|c|c|c|c|c|c|}
\hline
Br. No. & $V_s$ & $Rs$ & $R_{L}$ & $R_M$ & $\alpha$ & $V_D$ & $R_D$ & $I_{\min}$ & $V'_{\min}$ & $g_{\max}$ & $\lambda$ & $\mu$ & \color[HTML]{6200C9}$Is$& \color[HTML]{6200C9}$V'$& \color[HTML]{6200C9}$V''$& \color[HTML]{6200C9}$I$\\ \hline
1      & 45    & 0.5  & 0.04  & 0.022 & 0.0027   & 0.7   & 0.018 & 0.5364     & 14          & 4.3648     & 1         & 1.5 & \color[HTML]{6200C9} \begin{tabular}[c]{@{}c@{}}9.1187,\\9.118\end{tabular} & \color[HTML]{6200C9}\begin{tabular}[c]{@{}c@{}}40.4407,\\40.441\end{tabular}& \color[HTML]{6200C9}\begin{tabular}[c]{@{}c@{}}71.0097,\\71.007\end{tabular}& \color[HTML]{6200C9}\begin{tabular}[c]{@{}c@{}}5.0485,\\5.0469\end{tabular}  \\ \hline
2      & 50    & 0.4  & 0.053 & 0.025 & 0.0031   & 0.65  & 0.014 & 0.4244     & 13          & 3.9306     & 1.5       & 1  & \color[HTML]{6200C9}\begin{tabular}[c]{@{}c@{}}6.7501,\\6.7541\end{tabular} & \color[HTML]{6200C9}\begin{tabular}[c]{@{}c@{}}47.3,\\47.298\end{tabular}& \color[HTML]{6200C9}\begin{tabular}[c]{@{}c@{}}71.096,\\71.094\end{tabular}& \color[HTML]{6200C9}\begin{tabular}[c]{@{}c@{}}4.3842,\\4.3867\end{tabular}  \\ \hline
3      & 42    & 0.45 & 0.053 & 0.019 & 0.0029   & 0.6   & 0016  & 0.3226     & 12          & 4.0561     & 1         & 1   & \color[HTML]{6200C9}\begin{tabular}[c]{@{}c@{}}8.7935,\\8.793\end{tabular} & \color[HTML]{6200C9}\begin{tabular}[c]{@{}c@{}}38.0429,\\38.043\end{tabular}& \color[HTML]{6200C9}\begin{tabular}[c]{@{}c@{}}71.0505,\\71.047\end{tabular}& \color[HTML]{6200C9}\begin{tabular}[c]{@{}c@{}}4.5674,\\4.5659\end{tabular} \\ \hline
\end{tabular}}
\label{tab:caseiii}
\vspace{-0.2cm}
\end{table*}

\subsection{The Convex Relaxation}
\noindent The non-convexity of the optimization problem \eqref{eqn:orig_opt_problem} is the primary challenge. In a series of steps it shall be shown that the a convex relaxation of the problem can be used to obtain the globally optimal solution. As a first step, let the load voltage $V_{\rm load}$ be fixed to a value $\boldsymbol{V_{\rm load}}$ within the desired range. Let the solution set in this case be given by $\boldsymbol{P}=\left\{\boldsymbol{V_1},\cdots,\boldsymbol{V_m},\boldsymbol{V'_1},\cdots,\boldsymbol{V'_m},\boldsymbol{V''_1},\cdots,\boldsymbol{V''_m},\boldsymbol{Is_1},\cdots,\boldsymbol{Is_m}\right.$, $\left.\boldsymbol{I_1},\cdots,\boldsymbol{I_m}\right\}$. Let the optimal cost be \textbf{C}. Now, suppose for some $\boldsymbol{\delta}> 0$, $\boldsymbol{V_{\rm load}}-\boldsymbol{\delta}$ is also within the desired range. Then it is obvious that the point $\boldsymbol{P_{\delta}}=$ $\left\{\boldsymbol{V_1},\cdots,\boldsymbol{V_m},\boldsymbol{V'_1},\cdots,\boldsymbol{V'_m},\boldsymbol{V''_1-\delta},\cdots,\boldsymbol{V''_m-\delta},\boldsymbol{Is_1},\right.$, $\left.\cdots,\boldsymbol{Is_m}, \boldsymbol{I_1},\cdots,\boldsymbol{I_m}\right\}$ is a feasible point for the optimization problem with the load voltage set to $\boldsymbol{V_{\rm load}}-\boldsymbol{\delta}$, the cost of which is also $\boldsymbol{C}$. Note that the assumption \textbf{A6} is necessary here to prove feasibility of $\boldsymbol{P_{\delta}}$. Therefore, the optimal cost with the load voltage set to $\boldsymbol{V_{\rm load}}-\boldsymbol{\delta}$ cannot be more than $\boldsymbol{C}$. All these arguments also hold good with the addition cost term given in \eqref{eqn:additive cost}, since the difference between the output voltages of the DCC's in $\boldsymbol{P_{\delta}}$ remain unchanged.  \textit{Thus, one can at once set the load voltage to the minimum value in the desired range. This fact shall be assumed in the analysis to follow.}\\\\
For the second step, note that the sixth set of constraints consists of  all non-convex constraints, as the functions $f_k$'s need not be affine. But, since $f_k(.)$ are concave, a standard convex relaxation would be to replace equality with less than or equal to. That is:
\begin{equation}
    V_k = f_k(Is_k) ~~\longmapsto~~ V_k \leq f_k(Is_k)
    \label{eqn:vi_constraint}
\end{equation}
This relaxation allows for the output voltage to be lower than what the VI-characteristic curves prescribe. Similarly, the seventh set of constraints also comprises all non-convex constraints, since those are quadratic and affine. A convex relaxation of these constraints can be achieved in the following way. With the assumption that the $f_k(.)$ is a PWL concave decreasing function (see \textbf{A1}), note that
\begin{align}
    V_k Is_k = f_k(Is_k)Is_k = \min_{j\in [1,P_k]} \left\{ \beta_{k,j}Is^2_k + \gamma_{k,j}Is_k\right\}.
\label{eqn:concavity_of_VI}
\end{align}
Note that the expression above is a concave function as $\beta_{k,j}$'s are non-positive. Secondly, the term on the RHS of a constraint is convex as long as the conditions \eqref{eqn:conditions} hold. Thus, the seventh set of constraints last is replaced with the following set of convex constraints
\begin{align}
    &\min_{j\in [1,P_k]} \left\{ \beta_{k,j}Is^2_k + \gamma_{k,j}Is_k\right\} \geq Is_k^2\left(R_{\rm eff, 1} -R_{\rm eff, 2}\right)\cdots \nonumber \\&\cdots + \left(I_k + Is_k\alpha_{\rm k}\right)^2\frac{R_k}{2}  + \left(Is_k - sI_k\right)^2\frac{\left|R_{M,k}-R_{D,k}\right|}{2} \cdots\nonumber \\&\cdots + \alpha_{\rm k} V_{\rm load}I_k  + V_{D,k}\left(I_k+\alpha_kIs_k\right)+ I_k^2R_{\rm eff, 3} 
      + V_{\rm load} I_k
    \label{eqn:power_constraint}
\end{align}
In essence, the above relaxation implies that the power generated by a power source is at least the sum total of the losses in the branch and the power delivered to the load by that branch. As mentioned earlier, the DCC's would operate in CCM if 
\begin{align}
    Is_k \geq Is_{k,\min}~\forall k.
\end{align}
To prove tightness of the convex relaxation, the above lower bound on source current will converted into an equivalent lower bound on output current. That lower bound on output current $I_{k,\min}$ can be obtained as the positive root of the equation
\begin{align}
        &\min_{j\in [1,P_k]} \left\{ \beta_{k,j}Is_{k,\min}^2 + \gamma_{k,j}Is_{k,\min}\right\} = \cdots \nonumber \\&..~ Is_{k,\min}^2\left(R_{\rm eff, 1} -R_{\rm eff, 2}\right) + \left(I_{k,\min} + Is_{k,\min}\alpha_{\rm k}\right)^2\frac{R_k}{2} ~..\nonumber \\&..~ + \left(Is_{k,\min} - sI_{k,\min}\right)^2\frac{\left|R_{M,k}-R_{D,k}\right|}{2} + \alpha_{\rm k} V_{\rm load}I_{k,\min} ~.. \nonumber \\&..~ + V_{D,k}\left(I_{k,\min}+\alpha_kIs_{k,\min}\right) + I_{k,\min}^2R_{\rm eff, 3} 
          + V_{\rm load} I_{k,\min}.
          \label{eqn:quadraticequation}
\end{align}
However, if such a root does not exist, there is no feasible solution to the optimization problem. With all the aforementioned modifications, one can see that the main optimization problem given in Figure \ref{fig:mainalgo} is obtained.\\\\
As the third step it will be shown that, at optimality (if it exists), the convex relaxations are tight. To begin with, consider the constraint in \eqref{eqn:power_constraint} previously mentioned. Suppose on the contrary, for some $k$, at the optimal point $\boldsymbol{P}=\left\{\boldsymbol{V_1},\cdots,\boldsymbol{V_m},\boldsymbol{V'_1},\cdots,\boldsymbol{V'_m},\boldsymbol{V''_1},\cdots,\boldsymbol{V''_m},\boldsymbol{Is_1},\cdots,\boldsymbol{Is_m}\right.$, $\left.\boldsymbol{I_1},\cdots,\boldsymbol{I_m}\right\}$,
\begin{align}
    &\min_{j\in [1,P_k]} \left\{ \beta_{k,j}Is^2_k + \gamma_{k,j}Is_k\right\} > Is_k^2\left(R_{\rm eff, 1} -R_{\rm eff, 2}\right)\cdots \nonumber \\&\cdots + \left(I_k + Is_k\alpha_{\rm k}\right)^2\frac{R_k}{2}  + \left(Is_k - sI_k\right)^2\frac{\left|R_{M,k}-R_{D,k}\right|}{2} \cdots\nonumber \\&\cdots + \alpha_{\rm k} V_{\rm load}I_k  + V_{D,k}\left(I_k+\alpha_kIs_k\right)+ I_k^2R_{\rm eff, 3} 
      + V_{\rm load} I_k
\end{align}
Then, due to the continuity of the functions on either side of the constraint, there exists a $\delta>0$ such that the point $P_{\delta} = \left\{\boldsymbol{V_1},\cdots,\boldsymbol{V_m},\cdots \left(\boldsymbol{V'_k+\delta Rs_k}\right) \cdots,\boldsymbol{V''_1},\cdots,\boldsymbol{V''_m},\right.$ $\left.\cdots \left(\boldsymbol{Is_k-\delta}\right)\cdots\right.$, $\left.\boldsymbol{I_1},\cdots,\boldsymbol{I_m}\right\}$ is a feasible point with a lower cost. The point $P_{\delta}$ is feasible because a reduction in $\boldsymbol{Is_k}$ only increases the range for possible $Vs_k$ (see \textbf{A1}), and increases $V_k'$, thereby only requiring a reduction in gain of the $k^{\rm th}$ converter. However, the gain cannot go below 1 due to assumption \textbf{A6}. The reduction in cost can be reasoned as follows. A reduction in $Is_k$ surely reduces all terms in the cost function, possibly except $\left(Is_k-I_k\right)^2\frac{R_{M,k}}{2}$. But then, in a lossy boost converter $Is_k>I_k$ and thus $\left(Is_k-I_k\right)^2\frac{R_{M,k}}{2}$ can only reduce with a reduction in $Is_k$. \textit{Thus, at optimality, the power constraint is tight.} Since in the arguments of $P_{\delta}$, the DCC output voltages remain unchanged, the result holds good even with the additive cost of \eqref{eqn:additive cost}.\\\\
Lastly, note that the convex constraints pertaining to the VI-characteristics (given in \eqref{eqn:vi_constraint}) need not be tight at optimality. In other words, the voltage at the output of the power sources need not be equal to the value prescribed by the VI-characteristics. Suppose, at optimality, for the $k^{\rm th}$ converter,
\begin{align}
    \boldsymbol{V_k} < f_k\left(\boldsymbol{Is_k}\right).
\end{align}
Then, in that case, $\boldsymbol{V_k}$ can be increased to $\boldsymbol{V_k} + \delta$ where $\delta = f_k\left(\boldsymbol{Is_k}\right) - \boldsymbol{V_k}$, $\boldsymbol{V'_k}$ to $\boldsymbol{V'_k} + \delta$ and the gain can be reduced to 
\begin{align}
    \frac{\boldsymbol{V''_k}}{\boldsymbol{V'_k} + f_k\left(\boldsymbol{Is_k}\right) - \boldsymbol{V_k}}
\end{align}
to maintain $\boldsymbol{V''_k}$. Note that the gain remains greater than 1  due to \textbf{A7} and all currents in the network also remain unchanged in this process. \textit{Thus, one can find an optimal solution which adheres to the VI-characteristics.} Again, since these arguments left the DCC output voltages unchanged, the result holds good with additive cost of \eqref{eqn:additive cost}.

\begin{figure}[t]
    \centering
    \includegraphics[width=3.5in]{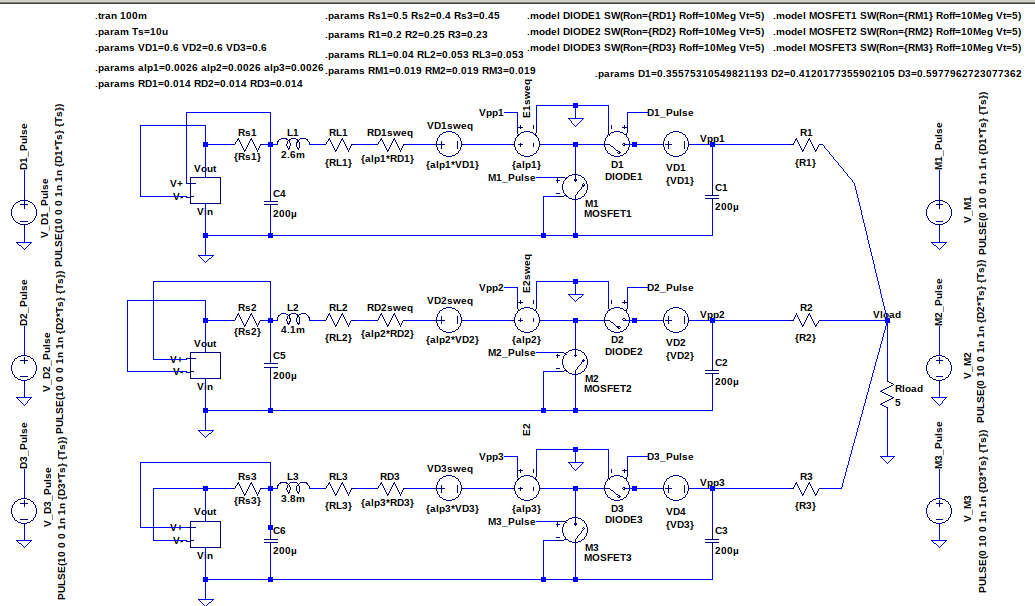}
    \caption{Case (i): An equivalent circuit for case (i). The non-linear voltage sources is realized using the circuit shown in Figure \ref{fig:case1_nlvs}.}
    \label{fig:case1_circuit}
\end{figure}

\begin{figure}[t]
    \centering
    \includegraphics[width=3.5in]{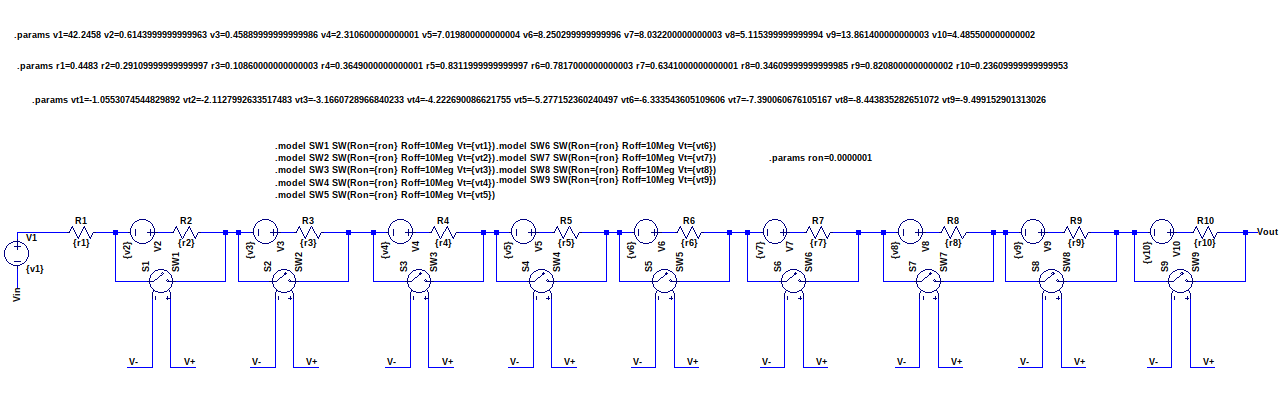}
    \caption{Case (i): The shown circuit realizes a concave and decreasing (w.r.t current) voltage source. With increasing current, a block of the voltage source with a resistance comes into effect, while it is bypassed until then. The increase in series voltage increases the intercept, while increase in series resistance increases the slope.}
    \label{fig:case1_nlvs}
\end{figure}

\begin{figure}[t]
    \centering
    \includegraphics[width=3.5in]{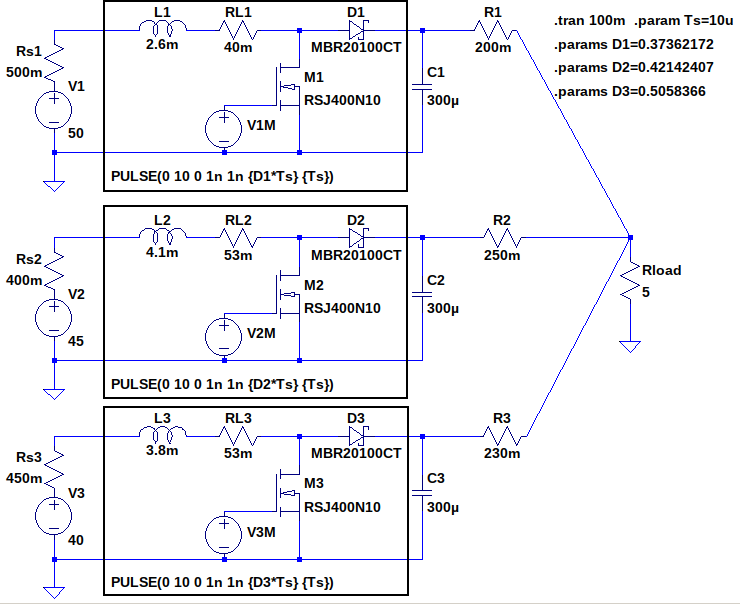}
    \caption{Case (ii, a): The LTSpice schematic shows three DCCs feeding power from three voltage sources with internal resistances to a resistive load. The semiconductor components (MOSFETs and diodes) are commercially available, and have been chosen appropriately as per the load voltage/current requirement. The simulation results for this circuit have been presented in Table \ref{tab:caseiia} and \ref{tab:caseiiapower}.}
    \label{fig:simulation_circuit_real_SC}
\end{figure}

\begin{figure}[t]
    \centering
    \includegraphics[width=3.5in]{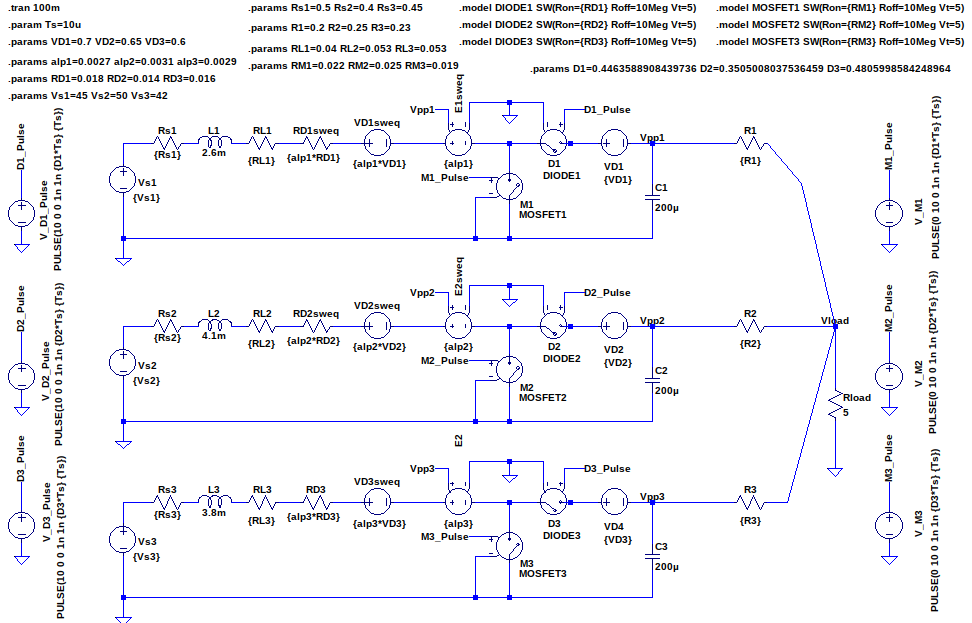}
    \caption{Case (ii, b): The LTSpice schematic shows equivalent circuits of three DCCs feeding power from three voltage sources with internal resistances to a resistive load. The semiconductor components (MOSFETs and diodes) are modeled as ideal switches with resistances/bias voltages, as applicable. The controlled voltage source aids in modeling the switching losses. The simulation results for this equivalent circuit have been presented in Table \ref{tab:caseiib}.}
    \label{fig:simulation_circuit_equivalent}
\end{figure}


\section{Computations and Simulations}
\begin{figure}
    \centering
    \includegraphics[width=3.5in]{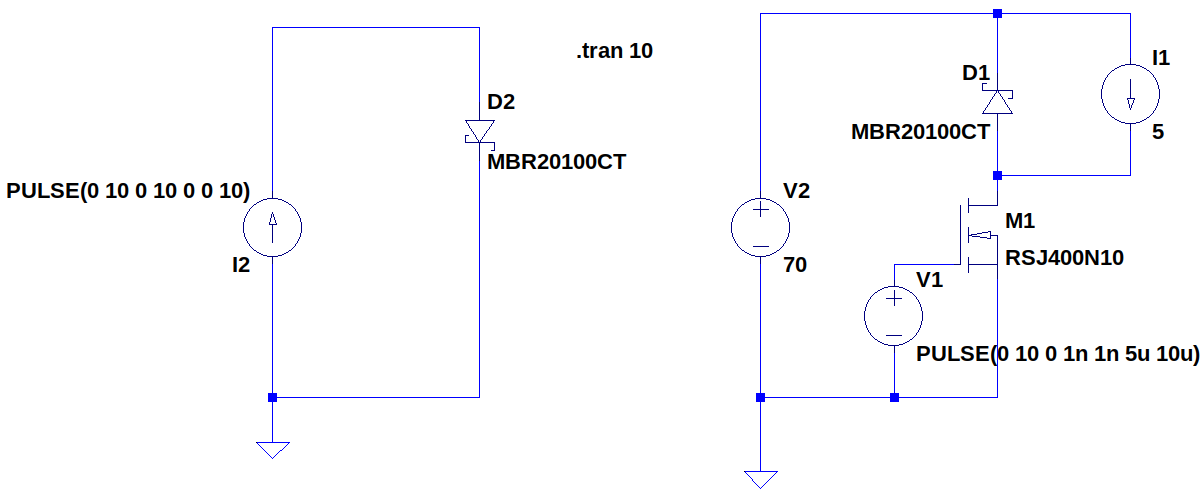}
    \caption{A schematic of the test circuits used to estimate the parameters pertaining to the diode and the MOSFETs.}
    \label{fig:test_circuits}
\end{figure}

\noindent This section presents computations in CVXPY and simulations on LTSpice. Three cases are presented: (i) the sources have non-linear VI-characteristics, (ii) the sources are constant voltage sources and the semiconductor characteristics are same across the DCCs, and (iii) the sources are constant voltage sources and the semiconductor characteristics are different across the DCCs. The optimal gains for all the cases are calculated using CVXPY and LTSpice simulations are presented. The LTSpice simulation for case (i) and (iii) consider an equivalent circuit which emulates the mathematical optimization model. The LTSpice simulations for case (ii) comprise two types: (a) simulation of a circuit with commercially available semiconductor components, and (b) simulation of an equivalent circuit, as mentioned earlier for other cases, which emulates the mathematical optimization model. In all the cases, the parameters of the circuit are assumed to be constant over the period of simulation. Table \ref{tab:caseisourceparams} present the parameters pertaining to the power sources for case (i), and Figure \ref{fig:case1_circuit} \& \ref{fig:case1_nlvs} present the circuits used for simulation on LTSpice. Table \ref{tab:caseiparamsandresults} presents the parameters pertaining to the other circuit elements and the numerical results from CVXPY \& LTSpice. Table \ref{tab:caseiia} and \ref{tab:caseiiapower} presents all the circuit parameters and results for case (ii,a). Similarly, the results for case (ii, b) are presented in Table \ref{tab:caseiib}. The circuits for  cases (ii, a) and (ii, b) are presented in Figure \ref{fig:simulation_circuit_real_SC} and Figure \ref{fig:simulation_circuit_equivalent}, respectively. The circuit parameters and results from CVXPY for case(iii) are presented in Table \ref{tab:caseiii}.\\\\
For case (ii), the diode forward bias and the resistance are estimated by first subjecting the diode to a slow voltage ramp up to the rated voltage (see the circuit to the left in Figure \ref{fig:test_circuits}). The current is recorded. The voltage and current are multiplied to get the instantaneous power consumed. The power curve is fitted with the form $P(I) = V_DI + I^2R_D$ using least squares, to obtain the forward bias voltage $V_D$ and the resistance $R_D$. The multiplicative constants for deriving switching losses are estimated using the circuit shown to the right in Figure \ref{fig:test_circuits}. The voltage source is set to 
\begin{align}
    V_{\rm load} + V_{D,k} + Is_{k,\min}\left(R_k+R_{D,k}\right),
\end{align}
since the voltage that would appear at the drain of the MOSFET in a DCC of the network would be at least this value. The current source is set to $\frac{V_{\rm load}}{3 R_{\rm load}}$, since the exact current sharing isn't known prior to optimization. The circuit in Figure 5 is then simulated for several cycles (setting duty ratio to 0.5), and the average power dissipated in the MOSFET $P_{\rm loss}$, and the rms-current through it $I_{\rm rms}$, are recorded. Since the drain-to-source resistance of the MOSFET $R_{M,k}$ can be read from its data-sheet, the multiplicative constant can be calculated as  
\begin{align}
    \alpha_k = \frac{P_{\rm loss} - 0.5\left(\frac{V_{\rm load}}{3 R_{\rm load}}\right)^2R_{M,k}}{\left(V_{\rm load} + V_{D,k} + \frac{V_{\rm load}}{3 R_{\rm load}}\left(R_k+R_{D,k}\right)\right)\frac{V_{\rm load}}{3 R_{\rm load}}}.
    \label{eqn:alphaestimation}
\end{align}
\newline The equivalent circuit emulating the losses modeled in the optimization formulation is shown in Figure 4. The terms of the loss appearing with squared of the source currents appear on the primary side of the switch. The terms appearing with squared of the output currents appear on the secondary side. The switch itself is an ideal switch with a small ON resistance equal to the drain-source resistance of the MOSFET. Since an ideal switch has no switching loss, the switching loss terms require a voltage controlled voltage source to be included in the model, essentially to emulate the product term of the source current and the output current. \\\\
As expected, the quadratic convex constraints are all tight at optimality, in all cases. in addition, the voltages and currents obtained from LTSpice are also in close correspondence with the numbers obtained from CVXPY. However, for case (ii, a) the difference between the two sets of numbers (obtained from CVXPY and LTSpice) is more as compared to other cases, which can be attributed to modeling error of the semiconductor components. 
\section{Future Work}
\noindent The convexity of the main optimization relies on certain conditions being satisfied by circuit parameters. Although, these conditions are often satisfied in practice, solving the otherwise non-convex problem is of considerable research interest. First is that the set of constraints given by \eqref{eqn:conditions}, which may not be satisfied in generality. Secondly, the condition for CCM operation of DCC's is chosen as \eqref{eqn:sourcecurrentCCM}. However, a necessary and sufficient condition would be
\begin{align}
    Is_k \geq \frac{V'_k}{2L_k f_s}\left(1-\frac{I_k}{Is_k}\right),~\forall k.
    \label{eqn:necandsuffCCM}
\end{align}
However, this is a non-convex constraint and makes the optimization problem harder.\\\\
The losses considered in the paper are not comprehensive. For example, the core losses in the inductor, the loss in the diodes due to reverse recovery current, the losses in through the output capacitances of the MOSFETs have not been considered. Although these are typically small as compared to the losses considered, their inclusion makes the treatment closer to reality. This might also be an interesting direction for future research. \\\\
The stability aspects of such a interconnected system is also an important matter for consideration, which is presently out of the scope of this paper. Although, simulations are indicative of a stable system, an analytical proof is desirable. This problem (seen in the sense of average models) in cases (ii) and (iii) boils down to finding checkable conditions for the stability of the set of matrices given by $\{u_1A_1+u_2A_2+\cdots+u_nA_n | 0\leq u_1,u_2,\cdots,u_n\leq 1 \}$. In case (i), the problem becomes harder due to the involvement of nonlinear voltage sources. \\\\ 
The presented methodology requires a communication between a central controller, which measures the parameters and runs the optimization routine, and the DCC's which actuate as per the set point communicated. Thus, a study along the lines of decentralized control might be of much importance. It might also be of interest to extend the presented approach to similar load-balancing systems with other converter classes, such as buck and  buck-boost. This would call for the losses in the DCC and the bounds on gains to be appropriately accounted for. 

\bibliographystyle{IEEEtran}
\bibliography{biblio}

\end{document}